\documentclass[conference,9pt]{IEEEtran}
\IEEEoverridecommandlockouts
\usepackage{cite}
\usepackage{amsmath,amssymb,amsfonts}
\usepackage{graphicx}
\usepackage{textcomp}
\usepackage{xcolor}
\usepackage{xspace}
\usepackage{booktabs}
\usepackage{makecell}
\usepackage{ragged2e}
\usepackage{algorithm}
\usepackage{algorithmicx}
\usepackage{mathtools}
\usepackage{pgfplots}
\usepackage[noend]{algpseudocode}
\begin{document}

\renewcommand{\Require}{\Statex \textbf{Input:} }
\renewcommand{\Ensure}{\Statex \textbf{Output:} }
\algrenewcommand{\algorithmiccomment}[1]{\hfill\textcolor{gray}{// \textit{#1}}}

\newcommand{\osort}{\textsc{\textbf{O-Sort}}\xspace}
\newcommand{\oscan}{\textsc{\textbf{O-Scan}}\xspace}
\newcommand{\otrans}{\textsc{\textbf{O-Trans}}\xspace}
\newcommand{\otransmerge}{\textsc{\textbf{O-Trans-Merge}}\xspace}
\newcommand{\osplittrans}{\textsc{\textbf{O-Split-Trans}}\xspace}
\newcommand{\ofilter}{\textsc{\textbf{O-Filter}}\xspace}
\newcommand{\omerge}{\textsc{\textbf{O-Merge}}\xspace}
\newcommand{\Null}{\textit{null}\xspace}

\algblockdefx[myindent]{Bindent}{Eindent}[1]{#1}{}
\makeatletter
\ifthenelse{\equal{\ALG@noend}{t}}%
  {\algtext*{Eindent}}
  {}%
\makeatother

\newcommand{\TODO}{\textcolor{red}{[TODO]}\xspace}
\newcommand{\todo}[1]{\textcolor{red}{[TODO: #1]}\xspace}
\newcommand{\euhost}{\emph{eu-2015-host}\xspace}
\newcommand{\SYS}{ObliGE\xspace}

\title{How Would Oblivious Memory Boost Graph Analytics on Trusted Processors?}

\author{%
    Jiping~Yu\IEEEauthorrefmark{1}\IEEEauthorrefmark{2}\IEEEauthorrefmark{3},
    Xiaowei~Zhu\IEEEauthorrefmark{2},
    Kun~Chen\IEEEauthorrefmark{2},
    Guanyu~Feng\IEEEauthorrefmark{1},
    Yunyi~Chen\IEEEauthorrefmark{1}\IEEEauthorrefmark{2}\IEEEauthorrefmark{3},
    Xiaoyu~Fan\IEEEauthorrefmark{1}\IEEEauthorrefmark{2}\IEEEauthorrefmark{3},
    and Wenguang~Chen\IEEEauthorrefmark{1}\IEEEauthorrefmark{2} %
    \IEEEcompsocitemizethanks{%
        \IEEEcompsocthanksitem\IEEEauthorrefmark{1} Tsinghua University.
        \IEEEcompsocthanksitem\IEEEauthorrefmark{2} Ant Group.
        \IEEEcompsocthanksitem\IEEEauthorrefmark{3} Work was done during the research internship at Ant Group.
    }%
}

\maketitle

\begin{abstract}
Trusted processors provide a way to perform joint computations while preserving data privacy. To overcome the performance degradation caused by data-oblivious algorithms to prevent information leakage, we explore the benefits of oblivious memory (OM) integrated in processors, to which the accesses are unobservable by adversaries. We focus on graph analytics, an important application vulnerable to access-pattern attacks. With a co-design between storage structure and algorithms, our prototype system is 100$\times$ faster than baselines given an OM sized around the per-core cache which can be implemented on existing processors with negligible overhead. This gives insights into equipping trusted processors with OM.
\end{abstract}


\section{Introduction}

In many scenarios, multiple parties can benefit from participating in joint computations on their aggregate data.
For example, multiple banks may share their blocklists to identify frauds that can hardly be detected using their own data; different hospitals can share patient data to train a better machine learning model that helps in medical diagnosis.
A basic requirement is to ensure the data privacy while allowing the parties to compute the results. 
Recently, there is a trend to perform multi-party data processing with trusted execution environments (TEE)~\cite{costan2016intel}, which may largely reduce the computation overhead, compared with cryptographic tools such as
Homomorphic Encryption (HE)~\cite{gentry2009fully} and
Secure Multi-Party Computation (MPC) protocols~\cite{shamir1979share,yao1982protocols,lindell2020secure}.
Under this setting, multiple parties first agree on a joint data processing task whose implementation is open to all participants for review, and then outsource the actual computation to a third-party (e.g. cloud service provider) with trusted processors.

Although many TEE implementations support transparent encryption of in-memory data, accessed memory addresses are not protected at the hardware level, which leaves the opportunity for curious eavesdroppers on the memory controller to infer a surprisingly large amount of information~\cite{xu2015controlled,liu2015last,ohrimenko2015observing,brasser2017software,gotzfried2017cache,lee2020off}.
To this end, we need to make the program running in the TEE to be data-oblivious~\cite{ohrimenko2016oblivious,zheng2017opaque,eskandarian13oblidb}, so that only insensitive information such as the dataset size can be inferred from the observed memory trace.
Thus, the algorithms need to be redesigned and become slower than plaintext versions, which reduces the performance benefits of using TEEs for multi-party data processing.
To mitigate such performance loss, many oblivious algorithms~\cite{ohrimenko2015observing,zheng2017opaque,eskandarian13oblidb,son2021oblicheck} assume there exists an oblivious memory (OM), to which the access patterns are assumed to be invisible to the attackers. Based on OM, the algorithms can achieve much better performance than the versions not relying on OM.

In this paper, we extend the above idea and study multi-party graph analytics on TEEs equipped with OMs. The reasons to pay special attention to algorithms on graphs are as follows.
(1) Graph analytics can attain better effects if utilizing multi-source data.
For example, in financial fraud detection~\cite{qiu2018real}, direct information sharing among financial institutions is prohibited, but with multi-party graph analytics, they can rely on the collaborative transaction graph to detect hidden fraud transactions which are  untraceable with each own data.
(2) Graph analytics is vulnerable to access-pattern attacks.
For example, the operation that ``for each edge from $u$ to $v$, update $w_1(v) \leftarrow w_1(v) + w_2(u)$'' is a common routine of graph algorithms such as PageRank~\cite{page1999pagerank}.
If directly implemented in TEE, even though the vertex data $w_1(v)$ and $w_2(u)$ are protected as ciphertext, $u$ and $v$ could be inferred from the addresses of memory access operations (or at least in cacheline-level granularities), which leaks sensitive information about the graph structure.
(3) Graph analytics is costly, and it is hard to achieve data-obliviousness.
To prevent information leakage, existing solutions~\cite{nayak2015graphsc} use expensive operations like sorting, which brings a lot of performance overhead, thus calls for a specially designed method that can make full use of OM.
Actually, as many graph algorithms are iterative, it is beneficial to reorganize the graph structure so that each subsequent iteration is performed more efficiently, which leaves quite a lot of design opportunities.

We design a prototype system named \SYS (\underline{Obli}vious \underline{G}raph \underline{E}ngine). Based on TEE with OM, we adopt a grid structure, which makes possible a highly-efficient oblivious graph scanning method, completely eliminating sorting passes from computation iterations.
We also propose pre- and post-processing methods to make the whole processing pipeline oblivious.
Our evaluation shows \SYS outperforms oblivious sort-based baselines by up to two orders of magnitude, and largely closes the performance gap compared with non-oblivious methods, making it practical for processing large-scale graphs.
We are delighted to see that OM can have much larger effects on boosting graph analytics, than on other applications such as sorting.
We also discuss the OM size requirements and the feasibility to implement such OM on trusted processors. We hope the results inspire hardware designers to equip trusted processors with OM.

\section{Preliminaries}


\subsection{Multi-Party Graph Analytics on Trusted Processors}

Graph analytics generally refers to iterative computation on graphs.
The edges are immutable, while the vertex data may change across iterations.
For example, PR (PageRank~\cite{page1999pagerank}), BFS (breadth-first search) and WCC (weak-connected components) are classic graph analytics applications.
For multi-party scenarios, each party holds parts of vertices and edges, wishing to compute the results on the merged graph.
We formulate the task of multi-party graph analytics on trusted processors as follows.

\subsubsection{Roles}

We identify the roles of $p$ clients, which provides input graph data and receives the results, and a central server equipped with trusted processors, which solely helps the computation and is typically provided by a cloud service.

\subsubsection{Input}

Each client $i$ inputs a private graph $(V_i, E_i)$.
$V_i$ denotes its set of vertices, while each element of $E_i$ is a pair $(u,v)$ such that $u,v \in V_i$, denoting an edge from $u$ to $v$.
The parties agree with $t$, the number of iterations, as a public information.

\subsubsection{Output}

After the computation, each client $i$ receives the resulting weight $w^t(v)$ for each vertex $v \in V_i$.
The method of weight calculation depends on the graph analytics application.
For example, for the PageRank algorithm~\cite{page1999pagerank}, we have
$$\textstyle w^t(v)=(1-f)+f \cdot \sum_{(u,v) \in E} \frac{w^{t-1}(u)}{d(u)}; w^0(v)=1$$
where $f$ is a constant damping factor (typically $0.85$), $d(u)$ denotes the out-degree of $u$ and $E$ is the set of all edges ($E=\cup_{i=0}^{p-1} E_i$).
Typically, an application starts with the vertex weights of the previous iteration, denoted by $w^{t-1}(\cdot)$, examines all edges $E$, and computes the weights for the subsequent iteration, $w^t(\cdot)$.
After completion of all $T$ iterations, these weights form the final result.

\subsubsection{Threat Model and Security Assumptions}

We consider clients to be semi-honest, implying that every client is expected to adhere precisely to the protocol (e.g., provide correct input data) while being keen on examining the private information (i.e. $V_i$, $E_i$) of other clients from received messages.
We assume that all the software and hardware components of the central server, except the trusted processor, may be compromised by a semi-honest adversary, which models the possible behaviors of an honest-but-curious cloud service provider.
Similarly to other applications built on trusted processors~\cite{ohrimenko2016oblivious,eskandarian13oblidb}, defending traffic timing attacks is out of the scope of this work.

\subsection{Usage of Trusted Processors}

Trusted processors, for example, with Intel Software Guard Extensions (SGX), allow programs to define private regions of memory named enclaves, whose content is inaccessible to untrusted software including operating systems and hypervisors.
Thanks to specialized hardware instructions and components for encryption/decryption available on modern CPUs, processing inside the enclave can thus even match the performance of insecure plaintext computations.
For multi-party data processing tasks, the computation workload can be outsourced to a third party with TEE.
The parties send their inputs to the TEE, and the predefined computation program evaluates the results and sends back to parties.
Many applications are built on this setting, such as machine learning~\cite{ohrimenko2016oblivious} and databases~\cite{eskandarian13oblidb}.

With the security guarantees of TEE, the only remaining information leakage is the memory access pattern, so that our main focus remains on designing a data-obliviousness algorithm to ensure that such patterns do not contain sensitive information.\footnote{This is because, for example, considering processors that incorporate Intel SGX, we emphasize several security features that it offers, e.g. \emph{software isolation} prevents software components from inspecting information that is visible only to the enclave, \emph{remote attestation} ensures the presence of a secure data channel between TEE and each client~\cite{ohrimenko2016oblivious}, and \emph{memory encryption} prohibits the adversary from deducing information about the plaintext data from the access traces~\cite{costan2016intel} (but note that the accessed address is not protected).}
While SGX opens up a new option for multi-party data processing, directly porting existing code to the enclave could still be insecure and might leak critical information like data samples through memory access patterns
~\cite{xu2015controlled,liu2015last,ohrimenko2015observing,brasser2017software,gotzfried2017cache,lee2020off}.
In particular, since the parties have no control over the hardware owned by the third party, they cannot detect peripheral-based attacks which make leakage of access patterns even easier~\cite{costan2016intel}.
Therefore, there is a requirement for specially crafted data-oblivious algorithms to prevent information leakage through these addresses.

\subsection{Oblivious Algorithm and Oblivious Memory (OM)}

To defend information leakage through memory access patterns, many oblivious algorithms were proposed, transforming the patterns so that the access sequences expose no information about secret data.
Scanning and sorting are two of the most important oblivious data processing primitives~\cite{ohrimenko2016oblivious}, and bitonic sort~\cite{batcher1968sorting} is widely adopted as an oblivious sorting algorithm, which needs $O(n \log^2 n)$ operations. 
There is a line of work on accelerating oblivious algorithms by leveraging states that are not observable by the adversary~\cite{ohrimenko2015observing,zheng2017opaque,eskandarian13oblidb,son2021oblicheck}.
A limited amount of oblivious memory (OM) is assumed to be inaccessible to untrusted adversaries.
For sorting, if an OM of size $s$ is provided, the time is reduced to $O(n \log^2 \frac ns + n \log s)$, by switching to non-oblivious sort when the sorting size is below the OM size. Other examples include $O(\frac{n^2}{s})$ database select and $O(\frac{nm}{s})$ hash join~\cite{eskandarian13oblidb}, where $n$ and $m$ are table sizes.

We assume that the trusted processor is equipped with an oblivious memory (OM), which has the same parameters of capacity, latency, and bandwidth as its L2 cache.
Possible hardware implementations of this OM and the cost will be discussed in Section \ref{sec:hardware}.
Note that this is more realistic than the full enclave page cache (EPC) being OM because EPC cannot be implemented on-chip.
This makes our security guarantee stronger than previous work~\cite{zheng2017opaque,eskandarian13oblidb,mishra2018oblix,ohrimenko2015observing}.

\begin{algorithm}[t!]
\caption{Basic oblivious routines adopted by \SYS}\label{algorithm:oblivious-routines}
\begin{algorithmic}[1]
\small
\Function{\osort}{array $A$, comparator $C$}
    \State $O(n \log^2 \frac{n}{s} + n \log s)$ bitonic sort~\cite{batcher1968sorting}
\EndFunction
\Function{\otrans}{array $A$, func $F$}
    \State $B \gets []$
    \ForAll {$a \in A$}
        \State $B.\text{append}(F(a))$
    \EndFor
    \State \textbf{return} $B$
\EndFunction
\Function{\otransmerge}{arrays $A_0,\dots,A_{n-1}$, func $F$}
    \State $B \gets []$ 
    \For {$i=0..n$}
        \ForAll {$j=0.. A_i.\text{len}$}
            \State $B.\text{append}(F(A_i[j], i))$
        \EndFor
    \EndFor
    \State \textbf{return} $B$
\EndFunction
\Function{\omerge}{arrays $A_0,\dots,A_{n-1}$}
    \State \textbf{return} $\otransmerge(A_0,\dots,A_{n-1}, (a, i) \mapsto a)$
\EndFunction
\Function{\osplittrans}{array $A$, int $n$, func $F$, func $G$}
    \State $B_0,\dots,B_{n-1} \gets []$
    \State \osort $A$ by $F(a)$ \Comment{$0 \leq F(a) < n$} 
    \ForAll {$a \in A$}
        \State $B_{F(a)}.\text{append}(G(a))$
    \EndFor
    \State \textbf{return} $B_0,\dots,B_{n-1}$
\EndFunction
\Function{\ofilter}{array $A$, func $F$} \Comment{$F(a) \in \{0, 1\}$}
    \State \textbf{return} $\osplittrans(A, 2, F, a \mapsto a)[1]$
\EndFunction
\end{algorithmic}
\end{algorithm}

For clearer further demonstrations, we list some oblivious routines in Algorithm \ref{algorithm:oblivious-routines}.
All of them ensure a deterministic access pattern as long as the arrays appearing as parameters or return values are of public known sizes.
They are adopted as building blocks of \SYS.

\section{Software Design}

This section presents the software components of \SYS, encompassing the design of the graph storage, the scanning algorithm to compute each iteration, and the additional pre-processing and post-processing techniques performed before and after the iterations.


\begin{figure}[t!]
\centering
\includegraphics[scale=.9]{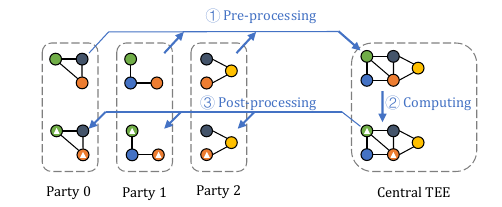}
\caption{\SYS workflow}
\label{overview}
\end{figure}

\SYS cooperatively runs among a central TEE (equipped with OM) and the participating clients of each party. As Fig. \ref{overview} shows, the main workflow of \SYS includes three steps: (1) pre-processing: the TEE gathers input information and merges the subgraphs into a large graph, (2) computing: the TEE performs the predefined algorithm on the merged graph and obtains the output results, and (3) post-processing: the TEE distributes to each client its output results (i.e. those that are associated with the party's known vertices).

\begin{table}[t!]
\centering
 \caption{Notations}
 \label{notations}
 \begin{tabular}{cl} 
 \toprule
 Symbol & Meaning (\# = ``the number of'') \\
 \hline
 $p$ & \# parties \\
 $n_i$ & \# vertices of party $i$'s subgraph, $|V_i|$ \\
 $m_i$ & \# edges of party $i$'s subgraph, $|E_i|$ \\
 $N$ & \# vertices of all subgraphs, $\sum_{i=0}^{p-1} n_i$ \\
 \hline
 $n$ & \# vertices of the merged graph ($n \leq N$) \\
 $m$ & \# edges of the merged graph ($m = \sum_{i=0}^{p-1} m_i$) \\
 $t$ & \# iterations \\
 $s$ & oblivious memory size \\
 \hline
 $k$ & \# vertices in each chunk \\
 $l$ & \# edges (including padding) in each block \\
 $l_i$ & \# edges (including padding) in each block of party $i$ \\
 \hline
 $b$ & \# chunks (thus, $b^2$ = \# blocks) \\
 \bottomrule
 \end{tabular}
\end{table}

Table \ref{notations} lists our notations, including the attributes of the subgraphs, the merged graph, and the computation. All (except $m_i$ and $m$) are defined to be public and are safe to be visible to all participants.

\subsection{Graph Storage Structure}

\begin{figure}[t!]
\centering
\includegraphics[scale=.9]{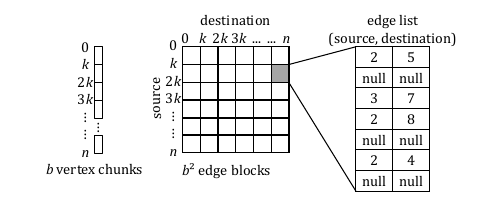}
\caption{Edge list structure}
\label{edgelist}
\end{figure}

\SYS basically supports a full-graph scan based on 2D grid-partitioned edge list structures, which is guaranteed to be completely data-oblivious with padding and does not expose sensitive information about the private data.
This 2D structure is typically employed in graph processing involving external disks~\cite{zhu2015gridgraph} to reduce I/O operations.
However, in our context, it serves an alternative purpose: ensuring that accesses reliant on sensitive data occur exclusively within the OM.
Vertex data are expressed as arrays.
Edge data are organized in a 2D grid structure with edges of each block.
As we assume the edges are immutable (while the vertex data may change across iterations), the storage structure can be static, without supporting fast modifications.
Thus, we use an edge list to store the edges from each block.
Fig. \ref{edgelist} shows the edge lists in a grid structure.
To build the structure, we first divide the $n$ vertices into chunks, each consisting of $k$ consecutive vertices, except for the last one that may contain fewer.
$k$ is chosen as the maximum value so that the data of $2k$ vertices can fit into the OM.
Assuming there are $b$ vertex chunks, we group the edges by the chunks which their source vertex and destination vertex respectively belong to, forming $b^2$ edge blocks.
For each block, we store a list of edges, where they are in arbitrary order.
The edge list includes additional null padding items, so that the edge lists from different blocks have the same length.

\subsection{Graph Scanning Algorithm}

\begin{algorithm}[t!]
\caption{Full-graph scan}\label{algorithm:full-graph-scan}
\begin{algorithmic}[1]
\small
\ForAll {$c_{dst} \in \text{Chunks}$ \textbf{in parallel}}
\Comment{$b$ times}
    \State Load vertex data of $c_{dst}$ into OM
    \Comment{$k$ vertices}
    \ForAll {$c_{src} \in \text{Chunks}$}
    \Comment{$b$ times}
        \State Load vertex data of $c_{src}$ into OM
        \Comment{$k$ vertices}
        \ForAll{$(src, dst) \in EdgeList(c_{src}, c_{dst})$}
            \If {not a padding edge}
            \Comment{$l$ edges}
                \State Computation on this edge
                \Comment{only accesses OM}
            \EndIf
        \EndFor
        \State Discard vertex data of $c_{src}$ from OM
    \EndFor
    \State Write back (and discard) vertex data of $c_{dst}$
    \Comment{$k$ vertices}
\EndFor
\end{algorithmic}
\end{algorithm}

\SYS provides an oblivious full-graph scan interface based on the 2D-grid edge list structure.
For various graph algorithms, different user-defined functions are specified, illustrating the operation for each edge.
Algorithm \ref{algorithm:full-graph-scan} shows the process of full-graph scan.
\SYS scans the 2D grid column by column, i.e. considering each destination chunk separately. 
For each column, we first load necessary data of destinations from the vertex array into the OM.
While scanning each block, the corresponding chunk of source vertex data is loaded into the OM, and then the edge list of the block is fully scanned, including the padding edges. 
After scanning over all edge lists of the column, the destination chunk of vertex data is written back from the OM to the outside, no matter whether the data is changed or not.
As the columns do not affect each other, we can parallelize the outer loop to utilize multiple CPU cores.

\textbf{Security Analysis.}
Consider a single-threaded version.
As comments show, everything except line 7 is oblivious because the memory accessing traces only depend on public parameters $b, k, l$. 
Line 7 is oblivious because the computation may only access the vertex data from either chunk $c_{src}$ or chunk $c_{dst}$, which we have already loaded into OM.
If we apply static-scheduled parallelization to the outer loop, the memory traces of each thread is also deterministic, and thus the multi-threaded version is also oblivious.
The OM usage per core is $O(1)$ plus two chunks of vertex data (loaded at lines 2 and 4), corresponding to the method of deciding $k$ in the previous section.

\subsection{Pre-Processing and Post-Processing}

\label{prepost}

Pre- and post-processing are critical parts to ensure that the whole application is oblivious.
The pre-processing includes vertex mapping and edge pre-processing, 
producing data structures used for computation.
In vertex mapping, we calculate the mapping relationship between the input vertices from the subgraphs and the vertices from the merged graph, and then get a merged list of original IDs of all vertices.
In edge mapping, we transform the edges specified by the original vertex ID into those specified by the mapped vertex ID, so that the edges can be used in the merged graph.
Then we partition the mapped edges into blocks, forming edge list structures.
The post-processing sends the results back to each party.
We carefully design 
efficient methods for the steps while keeping everything oblivious.

\subsubsection{Vertex Mapping}

\begin{algorithm}[t!]
\caption{Vertex mapping (performed by the central TEE)}\label{algorithm:sort-based-vertex-mapping}
\begin{algorithmic}[1]
\small
\ForAll {$i \in \text{Parties}$}
    \State Receive original Vertex ID $V_i$ from party $i$
    \Comment{$V_i\text{.len}=n_i$}
\EndFor
\State Init array $A = [\langle \text{OriginalID}, \text{Party}, \text{MappedID} \rangle]$
\State $A \gets \otransmerge(V_0, \dots, V_{p-1}, (v,i) \mapsto \langle v,i,\Null \rangle)$
\State \osort $A$ by OriginalID
\Comment{$A\text{.len}=N$}
\State $c \gets \Null;~~n \gets 0$
\Comment{current OriginalID and MappedID}
\Bindent {$A \gets$ \otrans $A$ with $\textbf{function}(a)$}
    \If {$c \neq a\text{.OriginalID}$}
        \State $c \gets a\text{.OriginalID};~~n \gets n + 1$
    \EndIf
    \State $a\text{.MappedID} \gets n$;~~\textbf{return} $a$
\Eindent
\State $c \gets \Null$
\Comment{current MappedID}
\Bindent {$A' \gets$ \otrans $A$ with $\textbf{function}(a)$}
\Comment{here $A'\text{.len}=N$}
    \If {$c \neq a\text{.MappedID}$}
        \State $c \gets a\text{.MappedID}$
    \Else
        \State $a\text{.MappedID} \gets \Null$
    \EndIf
    \State \textbf{return} $a$
\Eindent
\State $A' \gets \ofilter(A', a \mapsto a\text{.MappedID}\neq\Null)$
\Comment{$A'\text{.len}=n$}
\State $M_0,\dots,M_{p-1} = \osplittrans(A, p, a \mapsto a\text{.Party},$ \Statex \hfill $a \mapsto \langle a\text{.OriginalID}, a\text{.MappedID}\rangle)$
    \ForAll {$i \in \text{Parties}$}
        \State Send $M_i$ to party $i$
        \Comment{$M_i\text{.len}=n_i$}
    \EndFor

\end{algorithmic}
\end{algorithm}

To associate vertices of the same entity across multiple subgraphs, each vertex is specified by a unique key ID.
For example, if the key is phone number, and some vertices from different subgraphs share the same number, then the vertices should become one vertex in the merged graph.
The output is a mapping from the vertices of each subgraph to those of the merged graph, described by an array of (original ID, mapped ID) pairs for each party, as well as a global mapping array.
We assume the original IDs of all vertices are of the same length (otherwise, they can be padded into a large size) and are chaotic.
If not, we require all clients to do so (e.g. SHA-256 hash the IDs with a global salt randomly generated by parties), to ensure the load balancing among different blocks and to prevent other clients from obtaining knowledge about secret data (Section \ref{edgeprep}).
The mapped IDs are consecutive integers from 0 to $n-1$.

The idea of vertex mapping is based on oblivious sorting, illustrated in Algorithm \ref{algorithm:sort-based-vertex-mapping}.
We put together the original IDs on the central server, forming a merged list of (original ID, party ID, unknown mapped ID) tuples, and sort the merged list by the original ID (lines 1-5).
Then we can sequentially scan the merged list and assign mapped IDs (lines 6-10).
We remove consecutive equal mapped IDs to get the global result (lines 11-18). Then we split the results into each party's mapping, which are sent back for edge pre-processing (lines 19-21).


\textbf{Security Analysis.}
The vertex mapping is assembled from multiple oblivious routines. As in the comments, the lengths of $V_i,A,A',M_i$, are public. Thus, the whole post-processing is oblivious. The OM usage is $O(1)$ for temporary variables plus the buffer for \textsc{O-Sort}.

\subsubsection{Edge Pre-Processing} \label{edgeprep}

\begin{algorithm}[t!]
\caption{Edge pre-processing}\label{algorithm:edge-preprocess}
\begin{algorithmic}[1]
\small
\Bindent {At each client $i$......}
    \State Receive $M_i$ from central TEE
    \State $H_i \gets $ empty hash map
    \ForAll {$a \in M_i$}
        \State $H_i[a\text{.OriginalID}]=a\text{.mappedID}$
    \EndFor
    \State Init \textbf{grid} $G_i$, each block is edge list $G_i[x] = \{\langle \text{Src}, \text{Dst} \rangle \}$
    \ForAll {$e \in E$}
        \State $e' \gets \langle H_i[e\text{.OriginalSrcID}], H_i[e\text{.OriginalDstID}] \rangle$
        \State $G_i[\text{GetBlockID}(e')]\text{.append}(e')$
    \EndFor
    \State Resize the blocks so that all have equal size, filling null edges
    \State Send $G_i$ to central TEE
\Eindent
\Bindent {At central TEE......}
    \ForAll {$i \in $ Parties}
        \State Receive $G_i$ from party $i$
    \EndFor
    \State Init \textbf{grid} $G$, each block is edge list $G[x] = \{\langle \text{Src}, \text{Dst} \rangle \}$
    \For {$x \in 0..b^2$}
        \State $G[x] \gets \omerge(G_0[x], G_1[x], \dots, G_{p-1}[x])$
    \EndFor
\Eindent

\end{algorithmic}
\end{algorithm}

In edge mapping and partitioning, we transform the edges specified by original vertex IDs into those specified by mapped IDs, and dispatch them into a grid structure, forming an edge list for each block.
This step is performed distributedly, which means the clients also interactively participate in computing.



The main idea is to let each party help the central TEE to perform processing steps which are difficult to perform both efficiently and obliviously.
Thus, \SYS sends the mapping results back to each party at the end of the vertex mapping.
Algorithm \ref{algorithm:edge-preprocess} shows subsequent steps.
After each client receives its vertex mapping result, it locally transforms the edges into those specified by the mapped vertex IDs and groups the edges into $b^2$ blocks (lines 2-9).
The blocks are resized equally and sent to the central TEE (lines 10-11).
To reduce network traffic, each edge is sent as $2\lceil \log_2 k \rceil$ bits, specifying the offset of each mapped ID from the beginning of the chunk.
The central TEE only needs to merge the grids, forming a grid structure for the whole graph (lines 15-17).
The total complexity for each party $i$ is $O(n_i + m_i)$, and the complexity of the central TEE is $O(n+m)$.

\textbf{Security Analysis.}
Each client obtains the mapped vertex IDs of all vertices it knows.
For each vertex, the client knows the number of vertices from the merged graph which have smaller original IDs than it.
Thus, the clients can get some information like ``there are $x$ vertices with IDs between $y$ and $z$''.
As we assume the original vertex IDs to be safely obfuscated, the information received by each party should appear totally independently and identically distributed, and is useless to infer other parties' secrets because of the irreversibility of the obfuscating method.
From the server side, sending the mappings to clients and receiving the partitioned results from clients are oblivious because the data sizes are public parameters.

\subsubsection{Post-processing}

\begin{algorithm}[t!]
\caption{Post-processing}\label{algorithm:post-processing}
\begin{algorithmic}[1]
\small
\Require 
Global result, $R = [\langle \text{MappedID}, \text{Result} \rangle]$
\State Init array $S = [\langle \text{Party}, \text{OriginalID}, \text{MappedID}, \text{Result} \rangle]$
\State $Sp \gets \otransmerge(M_0,\dots,M_{p-1},$
\Statex \hfill $(v,i)\mapsto \langle i, v.\text{OriginalID}, v.\text{MappedID}, \Null \rangle)$
\State $Sg \gets \otrans(R,v \mapsto \langle \Null, \Null, v.\text{MappedID}, v.\text{Result} \rangle)$
\State $S \gets \omerge(Sp, Sg)$
\Comment{$Sp\text{.len}=N, Sg\text{.len}=n$}
\State \osort $S$ by (MappedID, Result == \Null)
\Comment{$S\text{.len}=N+n$}
\State $c \gets \Null$;~~$r \gets \Null$
\Comment{current MappedID and Result}
\Bindent{$S \gets$ \otrans $S$ with $\textbf{function}(s)$}
    \If {$c \neq s\text{.MappedID}$}
        \State $c \gets s\text{.MappedID}$;~~$r \gets s\text{.Result}$
    \EndIf
    \State $s\text{.Result} \gets r$;~~\textbf{return} $s$
    \Comment{assign Result to items}
\Eindent
\State $S \gets \ofilter(S, s \mapsto s\text{.Party} \neq \Null)$
\State $R_0, \dots, R_{p-1} \gets \osplittrans(S, p, s \mapsto s\text{.Party},$ \Statex \hfill \hfill \hfill $s \mapsto \langle s\text{.OriginalID}, s\text{.Result} \rangle)$
\Comment{$R_i\text{.len} = n_i$}
\ForAll {$i \in \text{Parties}$}
    \State Send $R_i$ to party $i$
\EndFor

\end{algorithmic}
\end{algorithm}

The post-processing mainly includes distributing the output data back to the parties.
Algorithm \ref{algorithm:post-processing} shows the method, given the results on the merged graph and the vertex mappings saved during vertex mapping.
First we merge the mappings and the results together (lines 1-4).
Then we sort the items by the mapped ID and scan them to assign the results (lines 5-10).
Fake elements are removed and the array is split into each party's result (lines 11-12).
Finally the results are sent back to the clients (lines 13-14).
Due to sorting, the complexity is $O(N \log^2 \frac{N}{s})$ with OM.

\textbf{Security Analysis.}
The post-processing is also assembled from multiple oblivious routines. As marked in the comments, the lengths of all arrays, $Sp, Sg, S, R_0, \dots, R_{p-1}$, are public parameters. Thus, the whole post-processing is oblivious. The OM usage is $O(1)$ for temporary variables plus the buffer for \textsc{O-Sort}.

\section{Discussion of Hardware Implementation}
\label{sec:hardware}

We argue that the expense of incorporating an OM with equivalent parameters (such as capacity, bandwidth, and latency) as the L2 cache is insignificant.
To substantiate this assertion, we outline two potential approaches to obtain the required OM functionality given the current capabilities of an existing Intel CPU with SGX, and show that the cost is negligible.
The two approaches are listed below.
\begin{itemize}
    \item \textbf{The addressable approach}: Use special instructions to turn the L2 cache into an addressable memory.
    \item \textbf{The pinning approach}: Allow pinning an address range onto L2 cache, which cannot be evicted until unpinning.
\end{itemize}

\subsection{The Addressable Approach}

In this approach, we expect to use special instructions to turn the L2 cache into an addressable memory, so that the accesses to these addresses are unobservable by the adversary, effectively making it an OM.
Actually, this technique already exists, known as Cache As RAM (CAR)~\cite{guide2011intel,amdcar}.
The most common usage of CAR is in the BIOS code, especially when the memory controller is not yet initialized~\cite{nallusamy2005framework}.
This is also used to protect sensitive data from the firmware~\cite{weis2014protecting} (note that their work cannot be adopted to our setting because we assume that the kernel may be compromised).

CAR currently cannot be enabled in trusted enclaves due to the lack of privilege to modify the corresponding control registers.
To realize this addressable approach, the processor can be augmented to allow such modification instructions to enable CAR inside the enclave.
To implement this, the processor does not need extra states.
Thus, this approach requires \emph{zero} additional bytes of SRAM (L2 cache) and even \emph{zero} additional flip-flops (registers), but only the wires and logic necessary to enable such modifications.
Since the logic is as simple as demonstrated above, this involves a negligible overall logic overhead.

\subsection{The Pinning Approach}

If we can pin a specific address range onto the L2 cache, we effectively make sure that any such accesses do not issue memory accesses and thus cannot be observed by the adversary.
The need to pin specific content to the L2 cache is similar to cache allocation technology (CAT)~\cite{selfa2017application,kim2019application}, which can control the affinity of cache ways.
Both SGX and CAT (for L2) are supported since the 4th generation Intel Xeon Scalable processors released in January 2023.
Currently, CAT supports assigning cache ways according to certain information such as the core number and/or the process identifier.

To implement this pinning method, it is necessary to adapt the above information to refer to the accessed (virtual) address.
This results in additional processor states that indicate whether the pinning method is activated, thus necessitating additional registers.
It also demands particular instructions, such as those that utilize model-specific registers.
However, this method does not require extra SRAM for the L2 cache nor alterations to each cache line because we can capitalize on the available space and capabilities of the current CAT setup.
The only requirements are an extra global register and wiring, which present less overhead compared to solutions that require additional SRAM storage or modifications for individual cache lines.


Comparing the two approaches, although the pinning approach involves greater overhead compared to the addressable approach, incorporating this feature offers increased flexibility, allowing the user to finely regulate the portion of L2 cache designated as OM, while maintaining the remaining space as standard L2 cache.
In contrast, the addressable method can only convert the entire L2 cache to OM.
Despite this, the pinning approach remains appealing as it does not require extra SRAM capacity or functionality.

Note that we did not employ simulator software like gem5 in this study because it primarily provides information on additional storage requirements (such as SRAM and registers) but fails to account for the overhead related to logic and wires, including chip area or power, since the logic is executed as software on the simulator host.
Our findings indicate that the addressable approach incurs no storage cost, whereas the pinning method only demands a state register to show if the feature is active.
Additionally, we did not pursue an FPGA-based experimental approach, as it would not accurately represent the performance of a contemporary CPU with tens of cores, each possessing an L2 cache at the MB level.
All our discussion about L2 cache does not apply to L3, because other CPU cores may easily pollute the state of the L3 cache (considering that the cloud service provider usually has other customers using the other cores), making the attacks easier to adopt and harder to detect.

\section{Evaluation}

Now we present various evaluation results for \SYS.
Since the hardware implementation cost has been discussed in the above Section \ref{sec:hardware}, here we focus on the methodology of software evaluation based on the assumed hardware.
To achieve this, we simulate the OM by specifying a memory region as large as the L2 cache and assuming that its accesses are invisible to the adversary.
This gives a lower bound of our performance results, i.e. our results are no better than that achieved by a real processor with OM (of the same bandwidth and latency with the L2 cache).
This is because our experimental results would match those of a real OM, provided that all data in the memory region were never evicted from the L2 cache.
In reality, evictions happen and the data may be reloaded, so our performance is not better than the real hardware.
Based on this methodology, we can conduct the experiments on an existing CPU.
In particular, we run the central TEE on a server equipped with an Intel Xeon Gold 6338 CPU, and all evaluated methods use its 32 CPU cores, each with 1.25 MB L2 cache.
Nevertheless, we will try different OM sizes in the experiments (Section \ref{omsize}), to demonstrate that \SYS can provide a consistent speedup for various OM sizes.

We mainly use \euhost graph~\cite{BMSB} for experiments.
It consists of 11.2 million vertices and 387 million edges.
Each vertex represents a domain host, and each directed edge indicates that there exists a link from
the source host to the destination.
We choose it because it is a publicly available webgraph and the vertices can be naturally grouped by their top-level domains as a ground-truth, corresponding to a multi-party scenario.
Each party holds the outgoing edges of its vertices.
The graph scale is suitable for performing processing on a single SGX-enabled CPU.
Conforming to our assumptions, we transform the original vertex IDs from domain strings to chaotic IDs of a fixed size, by hashing the string content and performing the AES encryption with a randomly-generated key. 

\subsection{Comparison with Existing Methods}

\begin{table}[t!]
\centering
\caption{Comparison with existing methods}
\label{generalcomparison}
 \begin{tabular}{c|c|c|c|c} 
 \toprule
 Method & PR & BFS & WCC & Speedup \\
 \hline
 Ligra (insecure) & 1.024 & 0.070 & 0.451 & 1/11.6$\times$ \\
 \hline
 \textbf{Our work, \SYS} & \textbf{3.322} & \textbf{3.250} & \textbf{4.670} & 1$\times$ \\
 \hline
 Sort-scan, with OM & 612 & 754 & 710 & 186$\times$ \\
 \hline
 Sort-scan, without OM & 1829 & 2735 & 2003 & 584$\times$ \\
 \hline
 ORAM (less secure) & 21006 & 24994 & 31677 & 6909$\times$ \\
 \bottomrule
\end{tabular}

\end{table}

To illustrate the large performance improvement over previous oblivious graph-unaware computing methods, we compare the computation time of \SYS with baseline methods in Table \ref{generalcomparison}.

\subsubsection{Sort-scan}

We adopt GraphSC's sort-scan~\cite{nayak2015graphsc} onto SGX as a baseline.
With OM, the execution time of sorting is 612 seconds on PR ($t=10$, the same below), 754 seconds for BFS, and 710 seconds for WCC.
We also report the performance of the version without OM, as 1829 seconds for PR (2.15$\times$ slower than with OM), 2735 seconds for BFS (3.63$\times$), and 2003 seconds for WCC (2.82$\times$). 
The sorting baseline only has moderate (2.80$\times$ geometric mean) performance improvements with OM, as $s$ only acts as the denominator inside the log-factor in its complexity $O((n+m) \log^2 \frac{n+m}s)$.

\subsubsection{\SYS}

The execution time is 3.322 seconds for PR (184$\times$), 3.250 seconds for BFS (232$\times$), and 4.670 seconds for WCC (152$\times$), achieving an overall 186$\times$ (geometric mean) speedup over the sorting baseline.
Compared with sorting, \SYS can better utilize the OM.
The space utilization (the proportion of non-null edges) is 36.49\%, showing the padding does affect the complexity for real-world graphs.

\subsubsection{Oblivious RAM (ORAM)}

We implemented the Path-ORAM protocol\footnote{ORAM is a general technique to hide memory access patterns, usually adopted in client-server storage scenarios, and can also be applied to TEEs with OM~\cite{eskandarian13oblidb}, but with additional memory bandwidth overhead.}~\cite{stefanov2018path} and compare it with \SYS.
We assume the ORAM is non-recursive, which makes the OM requirement larger than other implementations, as one layer of ORAM is not enough to reduce the need from all data to smaller than the OM size.
We also relax the security assumption (only protecting the lower 12-bits of the accessing index) and parallelize the algorithm.
Nevertheless, this implementation needs 21006 seconds for PR, which is 24.7$\times$ slower than sorting with OM, or 6323$\times$ slower than \SYS.
Other applications also suffers from thousands of times slowdown.
As a general solution, ORAM requires minimal modifications to the algorithm, but results in a low speed performance.

\subsubsection{Insecure System}

For reference, we also run Ligra~\cite{shun2013ligra}, a non-oblivious optimized implementation running outside SGX.
The overall speedup (geometric mean) over \SYS is 11.6$\times$.
Ligra's PR outperforms \SYS by 3.24$\times$, mainly because the padding in \SYS's storage structures.
Other applications achieve more speedup (46.4$\times$ for BFS, 10.4$\times$ for WCC) because insecure implementations can ignore a set of vertices and edges in some iterations, while \SYS has to scan all vertices due to security requirements.
Nevertheless, we can observe that \SYS has largely bridged the performance gap between secure and insecure solutions.

In conclusion, we show that
\SYS significantly (a few orders-of-magnitude) outperforms previous oblivious data processing techniques such as sort-based and ORAM-based methods. 

\subsection{Impact of Graph Scale}

\begin{figure}[t!]
\centering
\includegraphics[scale=.55]{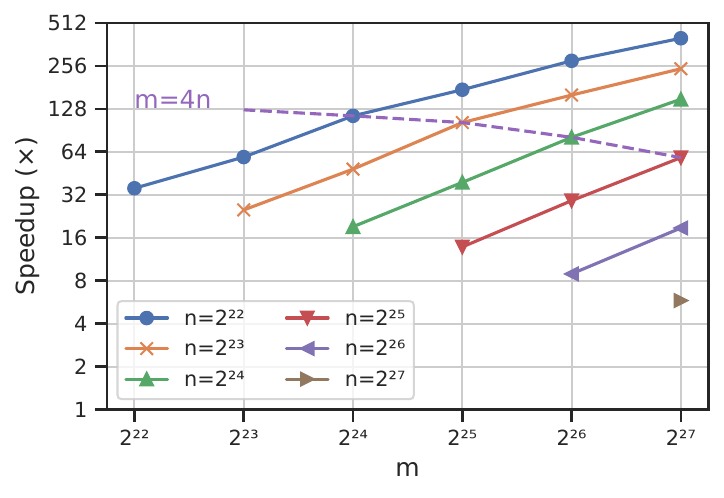}
\caption{Speedup of \SYS over sort-based computation for different scales}
\label{kron}
\end{figure}

To evaluate the speedup of \SYS over the sorting baseline (with OM) for different graph sizes, we generate a Kronecker graph~\cite{leskovec2010kronecker} for each pair of $n, m \in \{2^{22},2^{23},\dots,2^{27}\}$ and $n \leq m$.
Fig. \ref{kron} shows the speedup for the 21 generated graphs.
Each line represents a value of $n$ and the X-axis denotes $m$.
The speedup ranges from $5.8\times$ to $400\times$ for different graph scales.
If the average degree is at least 4 (on the dashed line or above), which is true for most real-world graphs, the speedup is at least $58\times$, showing a promising result.

\subsection{Impact of Oblivious Memory Size}

\label{omsize}

\begin{figure}[t!]
\begin{tikzpicture}
\begin{axis}[
    width=85mm,
    height=36mm,
    xmajorgrids=true,
    ymajorgrids=true,
    ymin=1,
    ymax=100000,
    xmin=0.5,
    xmax=5.5,
    ymode=log,
    xtick={1,2,3,4,5},
    xticklabels={$1/4 \times$,$1/2 \times$,$1 \times$,$2 \times$,$4 \times$},
    xlabel={\footnotesize OM size},
    ylabel={\footnotesize PR time (sec)},
    ytick={10,100,1000,10000},
    xtick style={draw=none},ytick style={draw=none},
    scaled x ticks=false,
    legend columns=-1,
    legend pos=north east,
    every tick label/.append style={font=\footnotesize},
]
\addplot[mark=square,brown,mark options=solid] coordinates { (1,582) (2,572) (3,612) (4,678) (5,752) };
\addplot[mark=x,blue] coordinates { (1,3.607) (2,3.431) (3,3.322) (4,3.893) (5,4.671) };
\legend{\footnotesize Sort-scan,\footnotesize \SYS}
\node[below] at (axis cs: 1,100) {\footnotesize 161$\times$};
\node[below] at (axis cs: 2,100) {\footnotesize 167$\times$};
\node[below] at (axis cs: 3,100) {\footnotesize 184$\times$};
\node[below] at (axis cs: 4,100) {\footnotesize 174$\times$};
\node[below] at (axis cs: 5,100) {\footnotesize 161$\times$};
\end{axis}
\end{tikzpicture}
\caption{PR time at different OM sizes}
\label{alphabeta}
\end{figure}
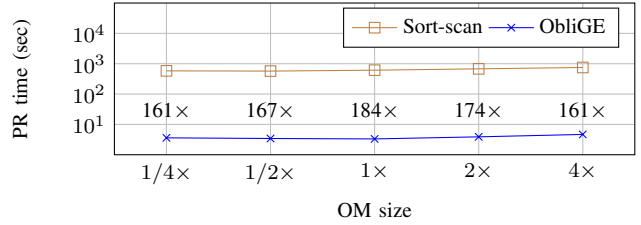

To evaluate the performance change given different amounts of OM sizes, we record the time of PR computation for \SYS and the sort-scan baseline on \euhost, shown in Fig. \ref{alphabeta}.
The sort-based approach benefits from larger OM sizes because more elements can use non-oblivious quicksort instead of oblivious implementation.
With a limited OM size, \SYS needs to split the graph into more blocks, resulting more padding edges and the vertex data is loaded for more times, effectively increasing the computation time.
The overall speedup of \SYS's computation over the sort-based method is over $160\times$ for OM size ranging from $1/4 \times$ to $4 \times$ the L2 cache size. Thus, \SYS has a consistent advantage for different OM settings.
  
\section{Related Work of Secure Graph Processing}

GraphSC~\cite{nayak2015graphsc} is a two-party graph analytics system built on garbled circuits~\cite{yao1982protocols}.
Its main idea is to put together the vertices and edges and use a sorting-scanning scheme.
Its complexity is $O((n+m) \log^2 (n+m))$ due to expensive sorting.
We adopt this idea and port it to TEE as a baseline.
Araki et al.~\cite{araki2021secure} propose a three-party solution based on secret sharing, which eliminates sortings from all iterations except the first, so that subsequent iterations only need linear time.
Graphiti~\cite{koti2024mathsf} is a following work based on a setting of two parties and a helper.
However, their techniques cannot be ported to centralized TEE scenarios because the TEE cannot securely act as multiple non-colluding parties in a data-oblivious manner.
GraphOS~\cite{chamani2024graphos} is a graph database on trusted processors.
It targets a mutable graph, while our work focuses on graph analytics assuming that the graph is immutable over iterations.
This results in a large difference regarding the execution speed, e.g. GraphOS needs $\sim 10^5$ seconds to process a graph of $2^{18}$ vertices and edges, while our work only needs $<10$ seconds to process a graph of $>2^{28}$ edges.

\section{Conclusion}

In this paper, we propose \SYS, a prototype system for multi-party graph analytics, based on trusted processors with oblivious memory (OM).
Our evaluation shows that \SYS can achieve a speedup of more than 100$\times$ over previous conventional sort-based implementations and largely closes the performance gap between secure and insecure solutions. 
The improvement is consistent with different sizes of graphs and various OM sizes ranging from $1/4 \times$ to $4 \times$ the L2 cache size.
The feasibility of implementing OM on trusted processors is also discussed.
We hope this inspires hardware designers to equip trusted processors with OM.

\bibliographystyle{IEEEtran}
\bibliography{main}
\end{document}